\documentclass[epj]{svjour}

\usepackage{graphics}
\usepackage{epsfig}
\usepackage{amssymb}
\usepackage{natbib}

\newcommand{\hx}{\hat{x}}
\newcommand{\hnu}{\hat{\nu}}
\begin{document}
\title{A Multi Agent Model for the Limit Order Book Dynamics}

\author{M. Bartolozzi\inst{1,2}}
\offprints{\\ mbartolo@physics.adelaide.edu.au}          
\institute{Research Group, Boronia Capital, Sydney NSW
2065 \and Australia Special Research Centre for the Subatomic
Structure of Matter (CSSM), University of Adelaide, Adelaide SA 5005}

\abstract{In the present work we introduce a novel multi-agent model with
the aim to reproduce the dynamics of a double auction market at microscopic time scale
through a faithful  simulation of the matching mechanics in the limit order book.
The agents follow a noise decision making process where their
actions are related to a stochastic variable,
{\em the market sentiment}, which we define as a mixture of public and private information.
The model, despite making just few basic assumptions over the trading strategies of the agents, is able to reproduce several
empirical features of the high-frequency dynamics of the market microstructure not only
related to the price movements but also to the deposition of the orders in the book.}
\PACS{
      {}{Market Microstructure}   \and
      {}{Econophysics}   \and
      {}{Detrended Fluctuation Analysis} \and
      {}{Multi-Agent Models}
} 
\maketitle

\section{Introduction}
\label{sec::introduction}

In the past few years, following the increasing power of technological infrastructures,
high-frequency trading, which broadly speaking includes every strategies which holding period is shorter than a day,
 has bloomed among the major financial institutions around the globe and, nowadays, it accounts for about 70\% of all the
 volume traded in US equities. The same trend is evident also in Europe where the number of transactions
 for the most liquid contracts has recently experienced an exponential growth:
 in the Eurex Stoxx futures index, for example, the number of daily trades has gone
from 12500 in January 2005 up to a maximum of 150000 in November 2008~\citep{Bartolozzi07b}.
As a consequence for the growing interest in the short time scales, the exchanges
 has started to provide live feeds of every single order submitted (buy or sell), therefore, making the market
 microstructure a primary field of interest for the financial practitioners.

The progressive shift towards the very high-frequencies has not been unnoticed in the Econophysics community which, in the
meanwhile, has become an established field of study among physicists~\citep{Bouchaud99,Mantegna99,Paul99,Voit05}.
In the multi-agent framework, a common feature in the physicist's approach to the market microstructure of {\em order driven markets} is the
  lack of a proper {\em utility function}, which is often invoked in Economics/Econometric literature in order to study the
  rational behaviour of the agents~\citep{Bailey05},  and, consequently, a simple statistical approach to the
  problem is preferred~\citep{Bak97,Maslov00,Matassini01,Raberto01,Smith03,Daniels03,Iori03,Farmer05,Mike08,Zaccaria09}.
This line of thought has been referred to as {\em zero intelligence}~\citep{Farmer05}: a
 recent review on some of the models proposed so far can be found in~\citep{Slanina08}. For an
  Econometric approach to the market microstructure, instead, the reader can refer to~\citep{OHara97}.

\begin{figure}
\vspace{1cm}
\centerline{\epsfig{figure=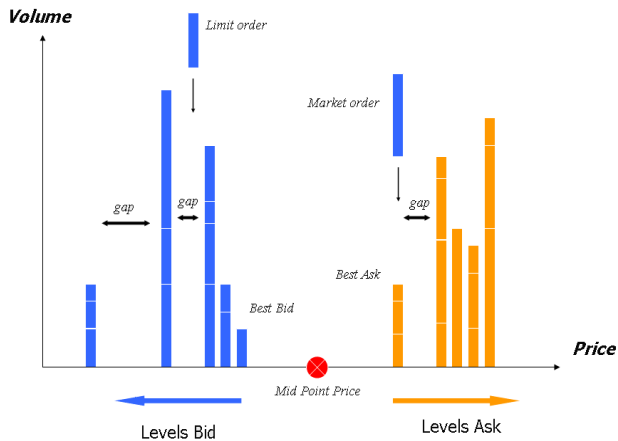,height=7cm,width=9cm}}
\caption{Cartoon representation of the LOB. Note that gaps between price levels can be present, especially
in non-liquid contracts.}
\label{fig::OrderBook}
\end{figure}

 Order driven markets are based on the principle of {\em continuous double auction} where the
  price of an asset is continuously adjusted in order to
account for the inflow of demand and supply. In particular, the orders placed by the
traders, that is the number of contracts they want to buy or sell along with the corresponding price, are
organized and matched in the {\em limit order book} (LOB) which ultimately defines the microstructure of the market.
At every instant in time, the LOB, a
  sketch of which is given for clarity in Fig.~\ref{fig::OrderBook}, is described by two sets of
{\em limit orders} in opposite ``directions", one for the long orders (buy side) and one for the short (sell side).
Each order is characterized by a {\em limit price}, that is the price the trader is willing to buy/sell, and
the number of contracts requested/offered, or volume. The minimum price difference greater than zero between two orders is
called {\em tick size} and, in real markets, depends on the specific contract.
When more than one limit order is send to the same price then the exchange rank them in a stack by arrival time:
the first to be executed will be the oldest\footnote{Note that while this ranking describes a typical situation,
 the execution of limit orders can slightly change depending on the rules of the specific exchange}.
 The {\em mid-point price}, $P_{m}$, which is usually referred
as the ``price" of an asset, is the mean value between the {\em
best ask} and the {\em best bid}, being, respectively, the
lowest price on the ask side and the higher price on the buy of the LOB. The execution of a limit order, that is the actual {\em trade}, is triggered if, and
only if, an order in the opposite direction matches its price quote.
 The incoming order can either be another limit order, in which case the order executed is at best ask or bid, or a
{\em market order}, that is a request to execute a certain volume, starting from the best price,
{\em immediately} and until it has been completely filled.
  This latter type of order is considered very ``aggressive"  given that the price of the spread is paid upfront and it
  is usually associated with ``impatience" traders.
 {\em Cancellations} also play an important role in the dynamics of the LOB. In fact, limit orders that  have not been filled
  in a reasonable amount of time, according to the trader's necessity, are usually removed from the LOB.

In the present work we introduce a novel multi-agent model\footnote{For ``agents" here we indicate all possible financial institutions such as investment banks, hedge funds etc...
rather that the occasional trader.} with the aim to reproduce the dynamics of the market microstructure.
Our agents relate their actions to a stochastic variable, the {\em ``market sentiment"},
which includes feedbacks from both public and private information. Besides, the model relies on a realistic LOB mechanics
which takes into account every possible change on ``event" base such as  the arrival of limit orders, market orders or cancelations
as well as a faithful order matching mechanism.

The paper is organize as following: in the next section we introduce our model while the
 numerical results of the simulations are presented in
Sec.~\ref{sec::Simulations}. Discussions and conclusion are left in the last section.

\section{The Model}
\label{sec::OrderBookModel}

\subsection{Brief outline}

Our market model evolves in discrete time steps\footnote{It is important to stress that in real markets the arrival of new information
 at microscopic time scale is heterogeneous in time and often people consider the {\em trade time} as a more appropriate scale.
While the output of the model is updated in discrete time steps for simplicity, the LOB is changed on event base.}
 during which each agent may undertake a certain action or just wait for a more profitable opportunity.  These actions
  can be broadly divided into {\em cancellation} and {\em active trading}, the latter including both limit and market orders.
 In particular, all decision steps are based on dynamical probabilities  which are function of both
 the private and public information, the former being related to the ``state" of the market as we will see in the following sections.
Moreover, we assume that our agents can have just one open position at the time and, therefore, before issuing a new order
they need to close their current position. By using this limitation we neglect explicit market making activity.

Another central part of our model is  the {\em order generation step} where the specifications for each order such as
{\em type} (limit or market), price (for limit orders) and volume  are decided.
In the next sections we explain in details all the fundamental building blocks.

\subsection{Public and private information update}

At the beginning of each time step, we assume that the agents have access to the current state of the LOB, that is they can ``see" every order placed in the
market: all the indicators derived by this knowledge, such as the mid-point price for example, are classified as {\em public information}.
However, these are not used in their ``raw" form but they get ``smoothed": this filtered information,
that we refer as ``perceived", is the one which is actually used in the process of decision making. The reason
behind this extra step is to take our simulations one step closer to reality where indicators are usually pre-processed  in
order to get rid of some noise\footnote{ As a consequence, short term memory is induced in the signal itself.}.
The filter used in our simulations is an exponential moving average (EMA)~\citep{Fusai08} which is defined for a
 generic time series $x(t)$ as
\begin{equation}
\label{eq::ema}
\hx (t) = \frac{1}{L} x (t)+\left( 1-\frac{1}{L}\right) \hx(t-1),
\end{equation}

where $L$ is proportional to the memory of the process.
 By using the previous equation Eq.(\ref{eq::ema}), we define the {\em perceived volatility} as
\begin{equation}
\label{eq::volatility}
\hnu(t) = \sqrt{\hat{r^{2}}(t)},
\end{equation}

being $r$ the one step return of the mid point price $r(t) = P_{m}(t) - P_{m}(t-1)$.
Note that, for time being, we have not included in the public information any  {\em news realize}, the importance of which has been
questioned over time, see ~\citep{Cutler89} for example or ~\citep{Joulin08} for a more recent criticism.

Regarding the {\em private information}, instead, we assume that it can be represented by a simple Gaussian process,
independent for each trader, with zero mean and standard deviation proportional to the perceived volatility of the market, $\hnu(t)$:
 this information can be thought to represent the convolution of the trading strategies used by the agent.

\subsection{Cancellation of orders form the LOB}

At each time step, agents having an outstanding order in the LOB will evaluate the possibility of
removing it according to two different criteria.
The first corresponds to a simple {\em time-out}, that is the order is automatically removed if it has not been executed in a  $T_{max}$
number of time steps relatively long if compared to the average time for a transaction
and fixed to 100 time steps in our simulations.
The second criterion, instead, is related to a strategic decision  based on the current market condition.
In particular, we define a {\em cancellation probablity}, $\psi_{\hnu}(t) \in [0,1]$, common to all the agents as
\begin{equation}
\label{eq::cancProbability}
\psi_{\hnu}(t) = 1 - {\rm e}^{-\gamma \, \hnu(t)},
\end{equation}
being $\gamma = 0.02$ a sensitivity parameter. The former expression, which
relies just on the perceived volatility $\hnu(t)$ in terms of information,  is justified
by the empirical observation that the cancellation frequency increases when the market
is highly volatile: this is clearly a case of self-reinforcing effect triggered by the herding behaviour of the
market participants~\citep{Cont00,Bartolozzi04}.

\subsection{Active trading and market sentiment}

While the cancellation step takes place, agents with no orders in the LOB evaluate
the possibility to enter the market.  Their decision is based on a
stochastic variable which represents the ``level of confidence" in their price forecast:
 the {\em market sentiment}, $\phi$. This quantity, which is a core part of our model,
 relates the public and the private information through a multiplicative process
and, specifically, for the $i$th agent at time $t$ we define
\begin{equation}
\label{eq::marketFeeling}
\phi_{i}(t) = \phi_{0}  \cdot  \kappa_{i}  \cdot \psi^{\star}_{\hnu} (t)  \cdot \eta (t) \cdot\epsilon_{i}(t),
\end{equation}
 where each terms represents a different aspect which may impact on the decision
 making process as  following:

 \begin{itemize}

  \item $\phi_{0}$ is a strength parameter common to all the agents which, as we will see in the next section,
   fixes the trading frequency of the model and, therefore, the time scale.

  \item The parameter $\kappa_{i}$ represents the agent's degree of {\em risk aversion}
   and it used in order to diversify the appetite for risk.
    Its value is randomly selected, at the beginning of the simulation,
     from a uniform distribution bounded in [0.25, 0.75].

   \item $\psi^{\star}_{\hnu} =  1 - \psi_{\hnu} (t) $ is the {\em volatility risk}  and mimics the fact
   that traders are more cautious to send orders during high volatility
   periods given that the risk associated to these is higher as well.

   \item $\eta (t)$, instead, is a proxy for the {\em liquidity risk}: a sparse LOB
   has a potential  large negative impact on the execution of a trade and, therefore,
   it decreases the probability that an agent is willing to accept the risk.
   In the present work we assume $\eta (t) = \bar{N}(t)/N$ where $N$ is the maximum number of orders in the LOB,
   equivalent also to the total number of agents in the simulation, and $\bar{N}(t)$ the number of orders in the LOB at time step $t$.
   From the definition we have that  $\eta (t) \in [0,1]$.

   \item $\epsilon_{i}(t)$ represents the {\em private information} which is drawn
   from a Gaussian distribution with zero mean and standard deviation equal to $\hnu (t)$,
   $\epsilon_{i}(t) = \hnu (t) \cdot G(0,1)$.

 \end{itemize}

The {\em market sentiment},  Eq.(\ref{eq::marketFeeling}), can be thought as the convolution between
the agent's trading strategies, the private information, and the risk factors evaluated
via the public information (volatility and liquidity in this case):
the stronger is the signal the more likely will be for the trader to take a decision.

The next step involves the mapping of  $\phi_{i}(t)$
 into a {\em trading probability}, $p_{i}(t) \in [0,1]$,  via a the following transfer function
\begin{equation}
\label{eq::orderProbability}
p_{i}(t) = \frac{2}{\pi} | {\rm arctg} \left[ \phi_{i}(t) \right] |,
\end{equation}
which represents the probability for the $i$th agent to submit an order at time step $t$.
 Conversely, an agent will not take any action with probability $1-p_{i}(t)$.

The {\em direction} of a trade, assuming that short selling is always allowed, $d_{i}(t) = +1$ for long orders (buy) and $d_{i}(t) = -1$ for
short orders (sell), is also derived from the market sentiment according to
\begin{equation}
\label{eq::marketDirection}
d_{i}(t)= \frac{ \phi_{i} (t)}{|\phi_{i} (t)|},
\end{equation}

as if $\phi_{i}(t)$ was a {\em momentum} indicator for the market direction. Besides
 it is worth underlying that, according to the previous definition, the direction of a trade depends only on
the sign of the private information, Eq.(\ref{eq::marketFeeling}),  and, therefore, there is 50\%
chances for a trade to be a buy or a sell\footnote{The market sentiment, $\phi_{i} (t)$, and consequently the trade direction,
 can be statistically skewed in one direction  depending on different market factors
and, therefore, leading to herding effects. The study of this phenomenon will be
the topic of future investigations.}.

\subsection{Order generation}
\label{sec::orderGeneration}

Each time an agent submits an order to the market its specifications are
defined in {\em order generation step} which addresses  the following two
points:

\begin{itemize}
    \item {\em Limit or market order?} The decision to submit a {\em limit} or a
    {\em market} order is related to the impatience
     of the trader: if an order needs to be filled as soon as possible
     he/she may accept to pay the cost of the spread upfront and send a market one.
    The alternative would be a less aggressive limit order that will execute just at a pre-fixed price:
     the further away from the opposite best the more time the order will require to get filled.

    \item {\em Size of the order?} The number of lots that the trader is willing
     to buy or sell can be related to several factors such as
     to the specific execution strategy, to the available liquidity in the
     market, the volatility at that point in time etc...
     Moreover, in a real trading environment single large orders are usually splitted into smaller ones in order to minimize
     the cost related to their impact. While the latter is an important issue in practice, in order to keep things simple,
      we do not tackle this problem in the present work: orders are sent in just one single chunk.
\end{itemize}

\begin{figure}
\vspace{1cm}
\centerline{\epsfig{figure=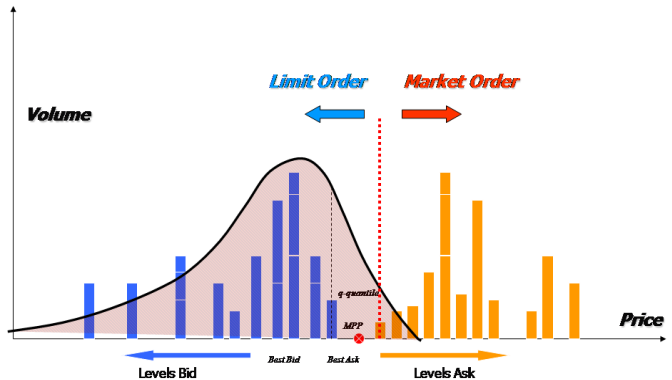,height=7cm,width=9cm}}
\caption{The price at which an order is submitted, and therefore
its ``aggressiveness", is drown from a log-normal distribution which
$q$-quantile is centered at the best bid (long order) or best ask
(short order). The ``direction" of the distribution is also chosen in accordance
with the order direction. In the cartoon above we sketch the probability of
price submission for a long (buy) order.}
\label{fig::OrderBookOrder}
\end{figure}

In our model the first point is addressed by setting the submission price of an order,
 which can be interpreted as the degree of ``aggressiveness", probabilistically via a value drawn from log-normal
 distribution\footnote{The parameters used in the simulations for the log-normal distribution are 7 for the mean and 10 for the standard deviation.},
 $\xi$, as following
\begin{equation}
\label{eq::priceLong1}
P_{l}= P_{b} - (\xi - Q_{P}),
\end{equation}

for a long order (buy) $P_{l}$ while for a short (sell), $P_{s}$

\begin{equation}
\label{eq::priceLong2}
P_{s}= P_{a} + (\xi - Q_{P}),
\end{equation}

 being $P_{b}$ and $P_{a}$ the best bid and the best ask, respectively, and
$Q_{P}$ the $q$-quantile,  $q=0.5$ in our case, of the distribution.  Besides,
 the value of $\xi$ is rounded up to an integer value and, therefore, fixing, without losing of generality, the tick size of
 our market  to 1.
 It is also important to stress that following the former procedure  we explicitly
assume that limit orders can be submitted asymptotically far from the best price as shown in the
 sketched in Fig.~\ref{fig::OrderBookOrder}.

After the submission price has been fixed, the {\em type} of the order is decided based on its relative
position to the best prices: if the submission price  results to be greater
than the ask price and the order is long (or lower than the bid price and the trade is short)
then we interpret this as a {\em market order} and all the volume will be completely filled starting from the best
price.
 All the other orders  are considered {\em limit orders} and, for a fixed price, they
  are organized in stacks from the oldest to the newest being the formers the first to be
  executed\footnote{ If the submission price coincides with the opposite best
 we give 50\% probability for the trade to be a limit order or a market order.}.
This modelling approach is motivated by two empirical observations:
the distribution of the volumes in the LOB, when averaged over long periods, displays
``fat tails"  and that the frequency of market order represents
just a fraction of that of limit orders submitted in the market~\citep{Bouchaud02}.
%

%

The second issue concerns the {\em order size}, that is the number
of contracts, or volume, that an agent is willing to buy or sell. Coherently with the stochastic
nature of this process, this is drawn from a
log-normal distribution\footnote{The parameters of the distribution
  used in the simulations are the same as for the price submission that is 7 for the mean and 10 for the standard deviation.},
  rounded to integer, with the following constrains: the order has
to be greater or equal than one and smaller than one quarter of the total volume contained in
the appropriate side of the LOB\footnote{In the simulation this limit is hardly reached:
 usually, single orders tend to be relatively small compared to the total volume available in the LOB.
 Moreover, before submitting a market order we also perform a liquidity check:
if the number of lots in the requested side of the LOB is less or equal $N_{min}$, with $N_{min} = {\rm min}(50,N/10)$,
 then the order is not issued. Liquidity checks on the volumes in the book are often done also in
  real trading in order to have an estimate of the risk associated with the trade itself.}.

Before moving to the next section, we wish to underline that the order generation step used in our
 model is just a first order approximation of what really happens in real execution strategies.
   In particular, we do not address the problem of splitting large volumes into smaller ones,
   typically used in order to minimize the transaction costs:
 this issue, despite being very important for market impact, goes beyond the scope of this work at present.

\section{Numerical simulations at short time scales}
\label{sec::Simulations}

\begin{figure}
\vspace{1cm}
\centerline{\epsfig{figure=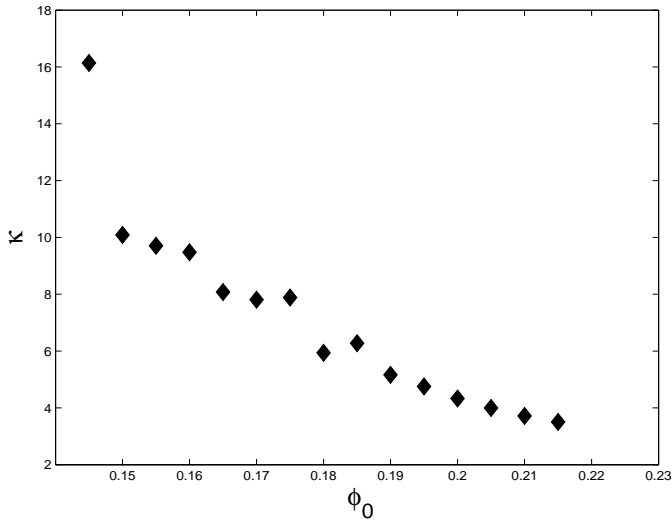,height=7cm,width=9cm}}
\caption{Kurtosis of the one step returns estimated over an ensemble of three simulations each one consisting in
$10^{5}$ time steps. It is noticeable how
this quantity converges to 3, the value for a Gaussian process, as $\phi_0$ increases. By decreasing $\phi_0$, instead, the activity
gets very low and the dynamics of the model starts to be ruled by few large market movements. The parameters for these simulation have
been declared in the previous section except for the EMA history $L$,  Eq.(\ref{eq::ema}), which in this instance has been fixed to 5.}
\label{fig::phi_kur}
\end{figure}

In this section we report the results of the numerical
 simulations of our microscopic market model for $N=10000$
 agents\footnote{ Numerical tests have shown that the outcomes are indifferent to the number of agents
 as long as their number is relatively large.}.
 The trading frequency, which can be interpreted as a time scale for the simulation,
  is fixed by the parameter $\phi_0$: as its value gets smaller  the trading activity becomes
 relatively lower and gaps within the LOB, leading to large price changes, are more likely to appear.
 On the other hand, if the activity is very high, the book is almost always full and large fluctuations will
 occur more sporadically. The scaling of the fluctuations with the activity is emphasized in
  Fig.~\ref{fig::phi_kur} where we report the kurtosis,
  $\kappa$, of the one step returns against $\phi_0$.

 For $\phi_0 = 0.165$, the value that we use in the rest of this section, the activity resembles that of the market at very short time scales,
 from seconds to minutes depending on the specific contract, where non-Gaussian fluctuations play a fundamental role.
 In this regime, which is the one of interest in the current work, there is a probability of approximately 30\%
 for the mid-point price to remain unchanged after one time step.

 Another important parameter is the EMA history, $L$ in Eq.(\ref{eq::ema}). In fact, we
  have found that activity clustering, such as high volatile periods, are particular evident when  $ 3 \lesssim   L  \lesssim  10$.
 In the simulations we have fixed $L=5$ for all the agents.

\subsection{Price and volatility dynamics}
\label{sec::priceDynamic}

\begin{figure}
\vspace{1cm}
\centerline{\epsfig{figure=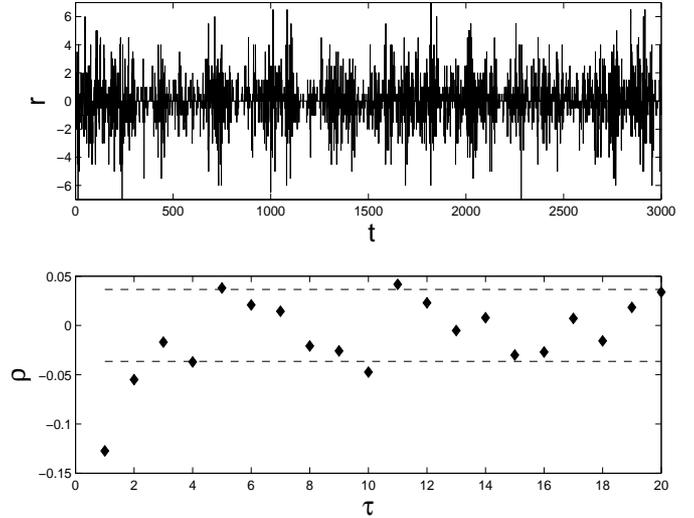,height=7cm,width=9cm}}
\caption{(Top) A window of the one step returns time series generated by the model with $N=10000$, $\phi_0 = 0.165$ and $L=5$. The time series,
standardized
according to $r(t) \rightarrow (r(t)- \bar{r})/\sigma(r)$ being $\bar{r}$ the average return and $\sigma(r)$ the standard deviation,
displays an intermittent character.
(Bottom) Autocorrelation function, $\rho(\tau)$, related to the time series on top: a significant
anticorrelation is evident for the first time steps. }
\label{fig::returns_corr}
\end{figure}

A sample of the time series of one step returns, $r(t)$,
 generated by the model  is reported in Fig.~\ref{fig::returns_corr}
 along with its autocorrelation function, $\rho(\tau)$. From the plots we can notice
  an intermittent dynamics, characteristic of financial
 time series at short time scales, as well as a significant negative correlation up to few time
steps. This latter effect, known empirically as ``bid-ask bounce",
in the present model is purely a result of the order book mechanics.
 However, for real high-frequency data this effect can last up to few minutes
  depending on the specific market~\citep{Bouchaud99}: this is an important indication that, on top
  of the LOB mechanics, there must be other mechanisms, such as memory feedbacks or order splitting for example, responsible for the
   enhancement of the negative correlation in price changes.

 The anticorrelated hehaviour of the returns is also confirmed by an estimate of the {\em Hurst exponent}, $H$,
 done via the {\em detrended fluctuation analysis} algorithm~\citep{Peng93,Bartolozzi07}.
 According to this method, originally developed during the Nile's river dam project \citep{Hurst51,Feder88},
  a time series is {\em persistent} if $H>0.5$, {\em  anti-persistent} if $H<0.5$ or {\em uncorrelated} if $H=0.5$.
The value found for the one step returns in our model is $H=0.495 (2)$ over an ensemble of 3 runs of $10^{5}$ samples each,
 being the error on the last digit, reported in the brackets, estimated via the bootstarp method~\citep{Efron94}. Noticeably,
 by removing the zero from the time series the value of the former exponent is statistically equivalent, in fact we find in this case
 $H_{0} = 0.491(3)$

\begin{figure}
\vspace{1cm}
\centerline{\epsfig{figure=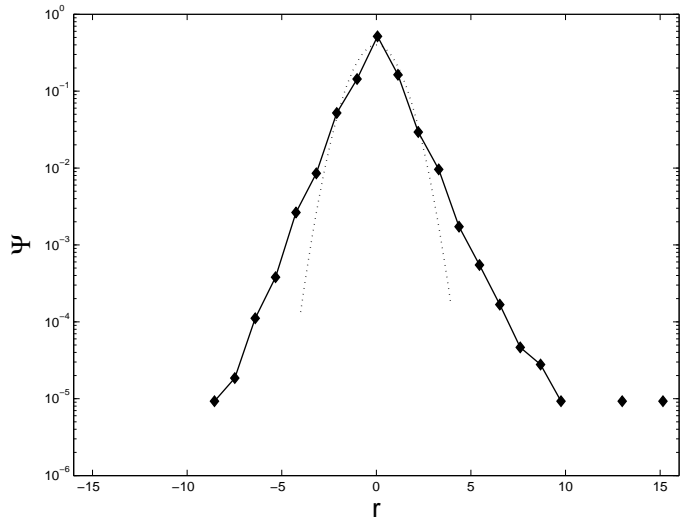,height=7cm,width=9cm}}
\caption{Probability distribution function, $\Psi$, for the one step returns standardized as in
Fig.~\ref{fig::returns_corr} (solid line). This distribution is {\em leptokurtic},
that is the tails are ``fatter" when compared to those of a Gaussian (dotted line). }
\label{fig::returns_pdf}
\end{figure}

The probability distribution function or {\em pdf}, denoted as $\Psi$, of the one step returns is reported in Fig.~\ref{fig::returns_pdf}
where the large fluctuations observed in Fig.~\ref{fig::returns_corr}
 (top) give rise to a {\em leptokurtic} shape of the distribution, that is the tails are ``fatter" than those of a Gaussian.
This feature, persistent up to time scales of weeks, is well documented in different empirical works~\citep{Bouchaud99,Mantegna99,Voit05}.
 However, while in most of the previous examples the asymptotic decay of the tails can be
described by a power law, in our simulations this is more consistent with an exponential one.
 From the same plot it is also possible to notice a relatively large presence of zero returns, which is also
 a  common feature of financial price time series at very high-frequencies.

\begin{figure}
\vspace{1cm}
\centerline{\epsfig{figure=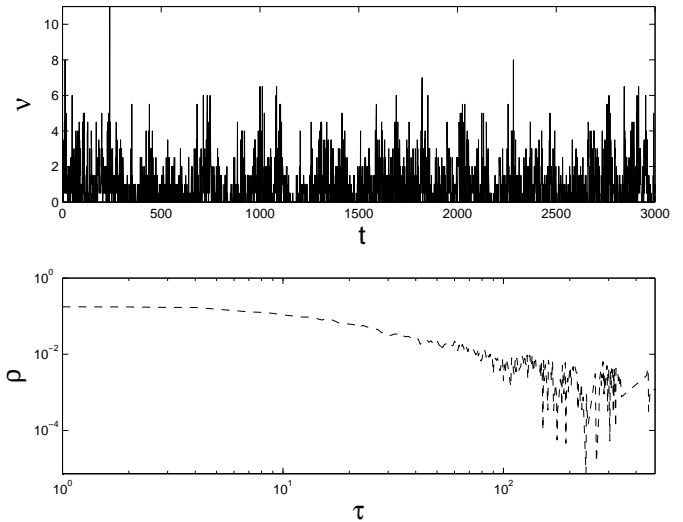,height=7cm,width=9cm}}
\caption{(Top) Absolute value of returns, or instantaneous volatility, for $N=10000$,  $\phi_0 = 0.165$ and $L=5$. Patches of high-volatility periods are visible.
(Bottom) Autocorrelation function relative to the time series on top.
The instantaneous volatility results to be correlated positively up to few hundreds steps.}
\label{fig::volatility}
\end{figure}

In Fig.~\ref{fig::volatility} (Top), instead, we report the {\em instantaneous volatility} defined as $\nu(t) = |r(t)|$ where
it is noticeable the presence of  {\em clustering}, that is periods of high activity tend to be {\em patched} together.
This memory effect is highlighted by the
slowly decaying autocorrelation function,  Fig.~\ref{fig::volatility} (Bottom), and by the Hurst exponent estimate
which gives the value of $H=0.61(2)$, indicating a significant persistency. Besides, we have verified that the correlation, despite being weaker,
 is still present after the zero returns removal, $H_{0}=0.56(1)$.
 A similar behaviour  is also observed in real markets, see, for example, in~\citep{Liu97,Gopikrishnan99,Liu99}.

As pointed out in the previous section, we would like to stress the role played by the
 memory induced via the {\em perceived volatility}, Eq.(\ref{eq::volatility}), and, therefore,
 the parameter $L$, in the volatility clustering.
  In fact, the clustering phenomenon seems to be particulary relevant only when the former
  feedback is present: this is a strong indication that the use
  of  short moving averages in high-frequency trading strategies, implicitly inducing memory effects, can be the source of
the bursts of patched  volatility observed at short time scales.
Also note that, following the previous observation, in our model both the market sentiment, Eq.(\ref{eq::marketFeeling}), and
the cancellation process, Eq.(\ref{eq::cancProbability}), contribute to the volatility clustering. In fact, it can be
shown via numerical experiments that by ``switching off"  the dependence from $\hnu(t)$ in one of
these two terms, for example by substituting it with a constant, the clustering effect gets noticeably reduced.

\subsection{Traded volume and impact}
\label{sec::tradedVolDynamic}

So far we have examined the bahaviour of the price returns and volatility. However, both these quantities are related to
 other fundamental factors, one of the most important being the {\em traded volume}, $V$.
 Some  authors, in fact, have recently argued that large transactions could be responsible for the non-Gaussian fluctuations
observed in the fat tails of the returns distribution~\citep{Gabaix03} even though
more recent empirical work seems to suggest that {\em liquidity crises}, that is  the temporary presence of gaps in the LOB,
should be at the origin of those~\citep{Bouchaud02,Daniels03,Farmer04,Lillo05,Weber05,Farmer06,Weber06}.
 Moreover, the dynamics of the traded volume time series, like the volatility, has been observed to be characterized
  by clusters of intense activity followed by relatively quite periods~\citep{Lillo04}.

\begin{figure}
\vspace{1cm}
\centerline{\epsfig{figure=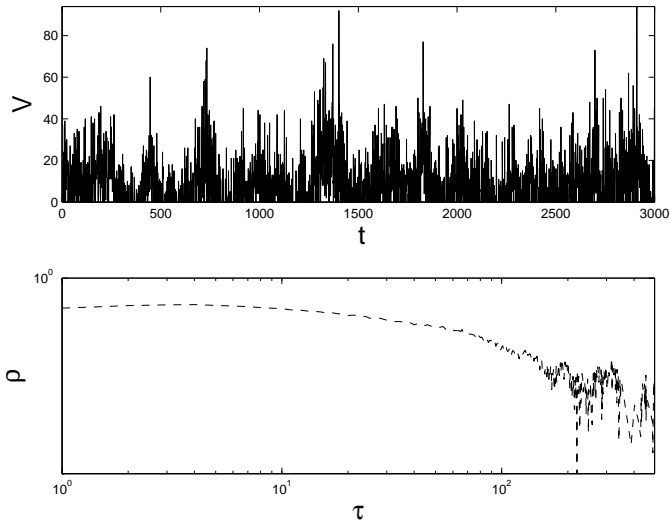,height=7cm,width=9cm}}
\caption{(Top) Window of 3000 samples for the traded volume in the model along with the relative autocorrelation function (Bottom).
The parameters used are $N=10000$,  $\phi_0 = 0.165$ and $L=5$.}
\label{fig::volume}
\end{figure}

   For our model, the cumulative traded volume during each time step is reported in Fig.~\ref{fig::volume}, where a behaviour
 very similar to that of volatility is evident. The intuitive  relation between the two is further confirmed by
 their cross-correlation coefficient which we found greater than 30\%.
 Moreover, the estimate Hurst exponent for the traded volume results
$H = 0.70(2)$ ($H_{0} = 0.66(1)$), matchings reasonably well the value found empirically in~\citep{Lillo04}, namely $H=0.73(7)$.
It also worth pointing out that a possible introduction of order splitting in the execution part of the model, which is missing at the moment,
 should enhance this correlation.
%

%

\begin{figure}
\vspace{1cm}
\centerline{\epsfig{figure=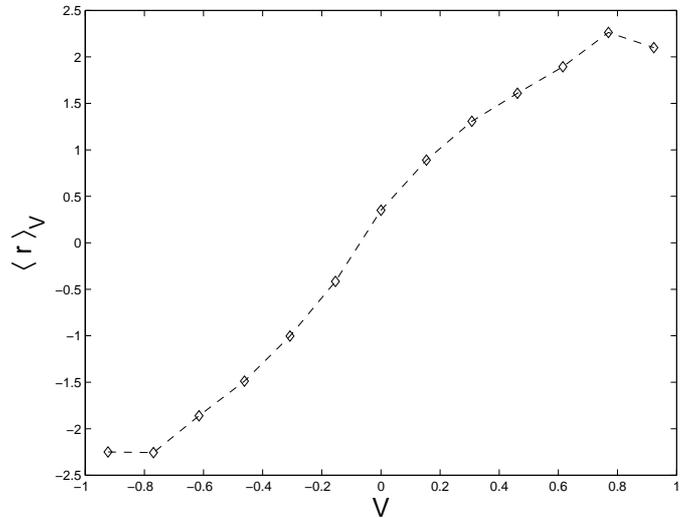,height=7cm,width=9cm}}
\caption{Instantaneous impact function where the negative part corresponds to sell initialized orders.
Note that while the impact is linear for $V \sim 0$, for large volumes this feature is lost.
Also, the values of the traded volume in the plot have been rescaled by the maximum volume traded in order to fit the interval [-1 1].
In this case the rescaling has been made simply to facilitate
the visual comparison of different impact functions in case of changing the parameters of the volume distribution, see Sec.~\ref{sec::orderGeneration}.
The simulation has been run with $N=10000$,  $\phi_0 = 0.165$ and $L=5$.}
\label{fig::impact}
\end{figure}

We also estimate the average {\em instantaneous } change in price subsequent to a trade of a certain volume $V$, usually known as
   {\em impact function}. This quantity is very important in practical applications being
   a proxy for the liquidity of the market: its value indicates how, approximately,  an order can penetrate ``deep" into the LOB.
The impact function for our model, Fig.~\ref{fig::impact}, displays a linear behaviour as long as $V$ is relatively small.
However, due to the presence of ``volume barriers", as we will see in the next section, this proportionality is lost as the volume increases. While the
 same shape for the impact function is observed in
real markets, see~\citep{Weber05,Weber06} for example, zero-intelligence models usually reproduce only the linear part, with the exception of~\citep{Farmer04}.

The values reported in Fig.~\ref{fig::impact} have also another important implication. In fact,
 the average price response to the large volumes is not enough to justify the
extreme fluctuations observed in the pdf of returns that can be greater than 10 ticks, Fig.~\ref{fig::returns_pdf}, which, therefore, have to be related or to
 rallies of orders in one direction
 or to  a temporary lack of liquidity between the levels of the LOB.
 Of course nothing prevents both scenarios to be realized at the same time.
%

\subsection{LOB average shape, spread and imbalance dynamics}
\label{sec::OBDynamic}

In this final section we turn our attention to the way the orders are ``on average" deployed in the LOB as well as
 the dynamics of some quantities related to the relative position of the orders such as the spread and the imbalance between buy and sell.

\begin{figure}
\vspace{1cm}
\centerline{\epsfig{figure=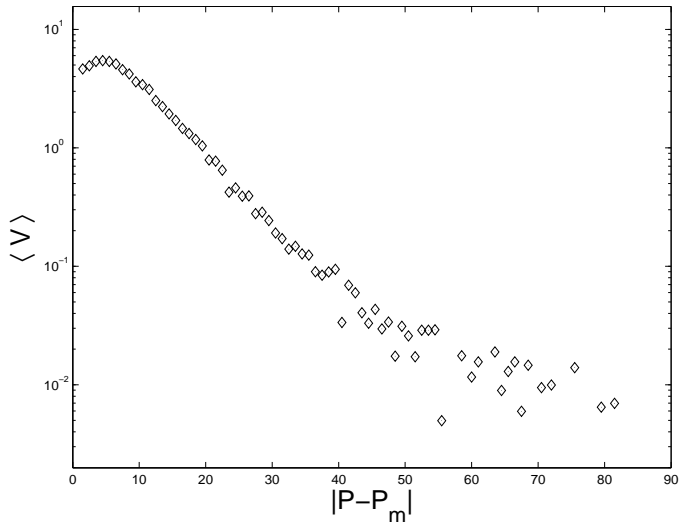,height=7cm,width=9cm}}
\caption{Average shape of the limit order book reported against the distance from the mid-point price.
  Note that, given the symmetry of the model, we have aggregated the data on the ask and the bid side of the book
   so to have  better statistics. The parameters used are the same as in Fig.~\ref{fig::volume}.}
\label{fig::avgBook}
\end{figure}

Firstly, we show the ``mean" shape of the LOB,
 that is the average volume present at a fixed distance, $|P-P_{m}|$, from
the mid-point price. From the plot,  Fig.~\ref{fig::avgBook}, it is clear that the liquidity increases steeply up
to a maximum, located few ticks away from $P_{m}$, and then it starts to decay as we move away
from it. The peak in volume can be thought as a sort of ``volume barrier" which prevents large orders to get too deep into the
LOB and, therefore,  confirming the speculations made in the previous section.
Very similar shapes have been found in empirical studies where it has been suggested that the asymptotic shape of the LOB,
related to the ``patient" traders, may follow a power law decay~\citep{Bouchaud02}.
It is also important to stress that the smooth ``mean" shape of the LOB is not significant
for the ``instantaneous" shape which, in fact, can be relatively sparse.
%

\begin{figure}
\vspace{1cm}
\centerline{\epsfig{figure=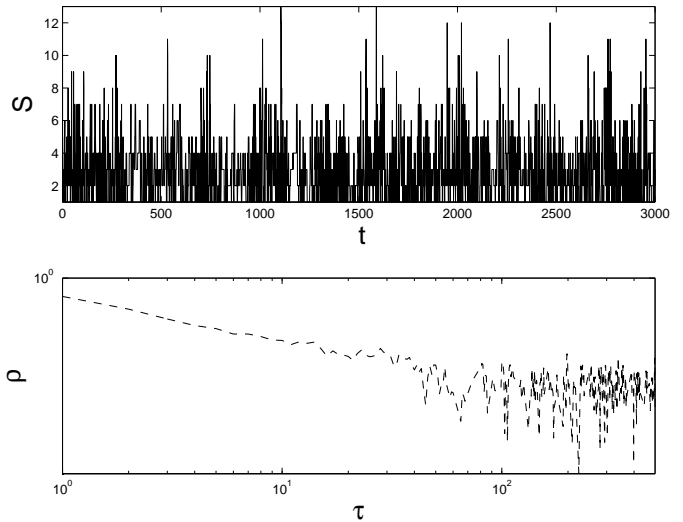,height=7cm,width=9cm}}
\caption{(Top) A sample for the spread time series (in ticks): it is possible to notice how, occasionally,
 the spread can become relatively large. (Bottom) Autocorrelation function for the spread time series.
 The parameters used are the same as in Fig.~\ref{fig::volume}.}
\label{fig::spread}
\end{figure}

Another important quantity related to the placement of the orders in the LOB is the spread, $S$. This is the difference
in price between the best ask and the best bid and it represents the
 cost that a trader has to pay upfront in order to execute a market order. Moreover, the spread
 is a further proxy for the liquidity:
large spreads would indicate shallow markets while very liquid ones will tend to keep a very small spread at all times,
possibly close to one tick.  Dynamically, the spread displays persistency in time with $0.7 \lesssim  H  \lesssim 0.8$ ~\citep{Plerou05,Cajueiro07,Gu07} depending on the market
and the time scale of observation, while the asymptotic shape of its pdf can be characterized by a power law function~\citep{Plerou05}.
The same two features appear also in our simulations, Fig.~\ref{fig::spread}, even though the Hurst exponent result to be relatively
smaller if compared to the empirical findings, namely $H=0.56(2)$ ($H_{0}=0.56(2)$):
this discrepancy may arise from the fact that at the moment we are considering only trivial execution algorithms.

\begin{figure}
\vspace{1cm}
\centerline{\epsfig{figure=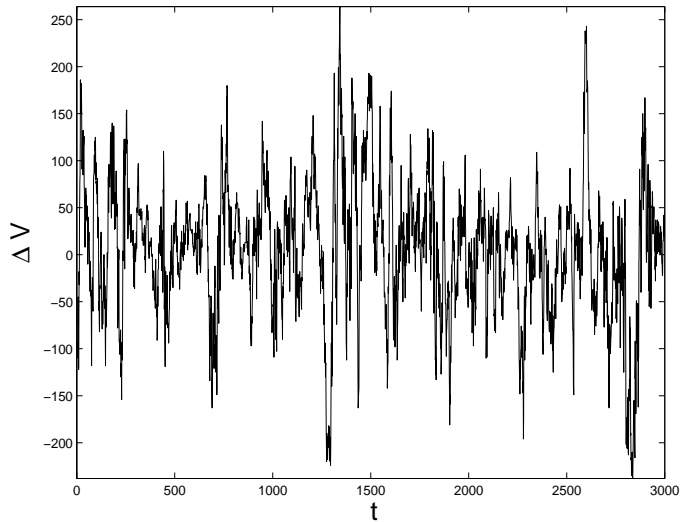,height=7cm,width=9cm}}
\caption{(Top) Time series of imbalance between demand and supply displaying a clear mean reverting nature.
The parameters used for the simulation are the same as in Fig.~\ref{fig::volume}.}
\label{fig::volImb}
\end{figure}

Lastly, we focus our attention to the dynamic imbalance between buy and sell orders or simply the {\em volume imbalance}, $\Delta V(t)$,
defined as
\begin{equation}
\label{eq::volImb}
\Delta V(t) = \sum_{i=1}^{N_{b}(t)}V_{i}^{b}(t) - \sum_{i=1}^{N_{a}(t)} V_{i}^{a}(t),
\end{equation}
where, for a time $t$, $V_{i}^{b,a}$ represents the volume of $i$th limit order and $N_{b,a}$ their total number,
respectively, for the bid and the ask side of the LOB.
The time series, Fig.~\ref{fig::volImb}, as expected for a market near-equilibrium,
displays a mean reverting behaviour, that is the imbalance tends to fluctuate all time, more or less symmetrically, around a
   ``pseudo-equilibrium" price, possibly close to criticality~\cite{Bartolozzi05,Bartolozzi06}. 
   The same dynamics has been observed for $\Delta V$ in different futures contracts~\citep{Bartolozzi07b}.
 %

\section{Discussion and conclusions}
\label{sec::conclusions}

In the present work we have developed a multi-agent framework characterized by a realistic order book keeping
 as a tool for the study of the activity of a double auction market at microscopic time scales.
 Inside this framework the model relies just on few basic assumptions
 related to the agent's strategic behaviour.
Among the most important ones, the order submission process makes use of a stochastic variable,
the {\em market sentiment}, which is related to both the public and the private information.
 Besides, the agents lack of the concept of ``utility maximization"  often invoked
  in the economics literature and embodying, in mathematical
 terms, the concept of ``rationality".

Despite its simplicity, the model manages to reproduce several empirical features of the high-frequency dynamics of the stock market
 such as the negative correlation in market returns, clustering in the trading activity
 (such as volatility, traded volume and bid-ask spread) as well as
 a non-linear response of the price change to the volume traded.
 Moreover,  the similarities with the real markets  extend also in the way the orders are deployed on
 the LOB as observed through the average shape of the book and the volume imbalance time series.

In conclusion, our model indicates that large part of the dynamics of the stock market at very short time scales can be
explained without the requiring any particular rational approach from an agent prospective if not some memory feedback
which, in our model, are represented by short term moving averages of the public information.
 Moreover,
our results confirm that large price movements are more likely to be related to a temporary lack of
 liquidity in the LOB rather than to large volume transactions.

Our future work will involve adding more realistic features and feedbacks in the model.
In particular, one interesting effect to take into consideration would be the {\em herding phenomenon} which,
it has been argued, to be at the origin of dramatic liquidity crises which can ultimately lead
 to financial crashes~\citep{Sornette04b}.
Other factors that can claim to play a fundamental role at these time scales are
 {\em news realize} and {\em market making strategies}: both of them should to be taken into account.

\section*{Acknowledgement}
The author would like to thank Prof. Tony Thomas and Prof. Derek Leinweber for a careful reading of the manuscript as well as
Matt Fender for the IT support.

\bibliographystyle{elsart-harv}
\bibliography{multiAgentLOB}

\begin{thebibliography}{45}
\expandafter\ifx\csname natexlab\endcsname\relax\def\natexlab#1{#1}\fi
\expandafter\ifx\csname url\endcsname\relax
  \def\url#1{\texttt{#1}}\fi
\expandafter\ifx\csname urlprefix\endcsname\relax\def\urlprefix{URL }\fi

\bibitem[{Bailey(2005)}]{Bailey05}
Bailey, R.~E., 2005. The economics of financial markets. Cambridge University
  Press, Cambridge, UK.

\bibitem[{Bak et~al.(1997)Bak, Paczuski, and Shubik}]{Bak97}
Bak, P., Paczuski, M., Shubik, M., 1997. Price variations in a stock market
  with many agents. Physica A 246, 430.

\bibitem[{Bartolozzi et~al.(2005)Bartolozzi, Leinweber, and
  Thomas}]{Bartolozzi05}
Bartolozzi, M., Leinweber, D.~B., Thomas, A.~W., 2005. Self-organized
  criticality and stock market dynamics: an empirical study. Physica A 350,
  451--465.

\bibitem[{Bartolozzi et~al.(2006)Bartolozzi, Leinweber, and
  Thomas}]{Bartolozzi06}
Bartolozzi, M., Leinweber, D.~B., Thomas, A.~W., 2006. Scale-free avalanche
  dynamics in the stock market. Physica A 370, 132--139.

\bibitem[{Bartolozzi et~al.(2007{\natexlab{a}})Bartolozzi, Mellen, and
  Chan}]{Bartolozzi07b}
Bartolozzi, M., Mellen, C., Chan, F., 2007{\natexlab{a}}. Internal report.
  Boronia Capital.

\bibitem[{Bartolozzi et~al.(2007{\natexlab{b}})Bartolozzi, Mellen, Di~Matteo,
  and Aste}]{Bartolozzi07}
Bartolozzi, M., Mellen, C., Di~Matteo, T., Aste, T., 2007{\natexlab{b}}.
  Multi-scale correlations in different futures markets. The European Physical
  Journal B 58(2), 207--220.

\bibitem[{Bartolozzi and Thomas(2004)}]{Bartolozzi04}
Bartolozzi, M., Thomas, A.~W., 2004. Stochastic cellular automata model for
  stock market dynamics. Physical Review E 69, 046112.

\bibitem[{Bouchaud et~al.(2002)Bouchaud, Mézard, and Potters}]{Bouchaud02}
Bouchaud, J.-P., Mézard, M., Potters, M., 2002. Statistical properties of stock
  order books: empirical results and models. Quantitative Finance 2(4), 251 --
  256.

\bibitem[{Bouchaud and Potters(1999)}]{Bouchaud99}
Bouchaud, J.-P., Potters, M., 1999. Theory of financial risk. Cambridge
  University Press, Cambridge.

\bibitem[{Cajueiro and Tabak(2007)}]{Cajueiro07}
Cajueiro, D., Tabak, B., 2007. Characterizing bid-ask prices in the brazilian
  equity market. Physica A 373, 627.

\bibitem[{Cont and Bouchaud(2000)}]{Cont00}
Cont, R., Bouchaud, J.-P., 2000. Herd behaviour and aggregate fluctuations in
  financial markets. Macroeconomics Dynamics 4, 170.

\bibitem[{Cutler et~al.(1989)Cutler, Poterba, and Summers}]{Cutler89}
Cutler, D., Poterba, J., Summers, L., 1989. What moves stock prices? Journal of
  Portfolio Managment 15, 4--12.

\bibitem[{Daniels et~al.(2003)Daniels, Farmer, Gillemot, Iori, and
  Smith}]{Daniels03}
Daniels, M.~G., Farmer, J.~D., Gillemot, L., Iori, G., Smith, E., 2003.
  Quantitative model of price diffusion and market friction based on trading as
  mechanistic random process. Physical Review Letters 90, 1008102.

\bibitem[{Efron and Tibshirani(1994)}]{Efron94}
Efron, B., Tibshirani, R., 1994. An Introduction to the Bootstrap. Chapman \&
  Hall, London.

\bibitem[{Farmer(2006)}]{Farmer06}
Farmer, J.~D., 2006. Comment on 'large stock price changes: volume or
  liquidity?'. Quantitative Finance 6(1), 1--3.

\bibitem[{Farmer et~al.(2004)Farmer, Gillemot, Lillo, Mike, and Sen}]{Farmer04}
Farmer, J.~D., Gillemot, L., Lillo, F., Mike, S., Sen, A., 2004. What really
  causes large price changes? Quantitative Finance 4(4), 383--397.

\bibitem[{Farmer et~al.(2005)Farmer, Patelli, and Zovko}]{Farmer05}
Farmer, J.~D., Patelli, P., Zovko, I.~I., 2005. The predictive power of zero
  intelligence in financial markets. Proceedings of the National Academy of
  Science pf the USA 102, 2254--2259.

\bibitem[{Feder(1988)}]{Feder88}
Feder, J., 1988. Fractals. Plenum Press, New York \& London.

\bibitem[{Fusai and Ronconi(2008)}]{Fusai08}
Fusai, G., Ronconi, A., 2008. Implementing models in quantitative finance:
  methods and cases. Springer-Verlag, Berlin.

\bibitem[{Gabaix et~al.(2003)Gabaix, Gopikrishnan, Pleru, and
  Stanley}]{Gabaix03}
Gabaix, X., Gopikrishnan, P., Pleru, V., Stanley, H.~E., 2003. A theory of
  power-law distributions in financial markets. Nature 423, 267.

\bibitem[{Gopikrishnan et~al.(1999)Gopikrishnan, Plerou, Amaral, Meyer, and
  Stanley}]{Gopikrishnan99}
Gopikrishnan, P., Plerou, V., Amaral, L., Meyer, M., Stanley, H.-E., 1999.
  Scaling of the distribution of fluctuations of financial market indices.
  Physical Review E 60, 5305.

\bibitem[{Gu et~al.(2007)Gu, Chen, and Zhou}]{Gu07}
Gu, G.-F., Chen, W., Zhou, W.-X., 2007. Quantiying the bid-ask spread in the
  chinese stock market using limit-order book data. European Physical Journal B
  57, 81 -- 87.

\bibitem[{Hurst(1951)}]{Hurst51}
Hurst, H., 1951. Long-term storage in reservoirs. Trans. Amer. Soc. Civil Eng.
  116, 770--799.

\bibitem[{Iori et~al.(2003)Iori, Daniels, Farmer, Gillemo, Krishnamurthy, and
  Smith}]{Iori03}
Iori, G., Daniels, M.~G., Farmer, J.~D., Gillemo, L., Krishnamurthy, S., Smith,
  E., 2003. An analysis of price impact function in order-driven markets.
  Physica A 324, 146.

\bibitem[{Joulin et~al.(2008)Joulin, Lefevre, Grunberg, and
  Bouchaud}]{Joulin08}
Joulin, A., Lefevre, A., Grunberg, D., Bouchaud, J.-P., 2008. Stock price
  jumps: news and volume play a minor role. Wilmott Magazine Sep/Oct, 1--7.

\bibitem[{Lillo and Farmer(2004)}]{Lillo04}
Lillo, F., Farmer, M. J.~D., 2004. Long memory of the efficient market.
  Nonlinear Dynamics and Econometrics 8(3), 1.

\bibitem[{Lillo and Farmer(2005)}]{Lillo05}
Lillo, F., Farmer, M. J.~D., 2005. The key role of liquidity fluctuations in
  determining the large price changes. Fluctuations and noise letters 5(2),
  L209.

\bibitem[{Liu et~al.(1997)Liu, Cizeau, Meyer, Peng, and Stanley}]{Liu97}
Liu, J., Cizeau, P., Meyer, M., Peng, C.-K., Stanley, H., 1997. Correlations in
  economic time series. Physica A 245, 437--440.

\bibitem[{Liu et~al.(1999)Liu, Gopikrishnan, Cizeau, Meyer, Peng, and
  Stanley}]{Liu99}
Liu, Y., Gopikrishnan, P., Cizeau, P., Meyer, M., Peng, C.-K., Stanley, H.,
  1999. Statistical properties of the volatility of price fluctuations.
  Physical Review E 60, 1390.

\bibitem[{Mantegna and Stanley(1999)}]{Mantegna99}
Mantegna, R.~N., Stanley, H.~E., 1999. An introduction to econophysics:
  correlation and complexity in finance. Cambridge University Press, Cambridge.

\bibitem[{Maslov(2000)}]{Maslov00}
Maslov, S., 2000. Simple model of a limit order-driven market. Physica A
  278(3-4), 571--578.

\bibitem[{Matassini and Franci(2001)}]{Matassini01}
Matassini, L., Franci, F., 2001. How traders enter the market through the book.
  preprint: cond-mat/0103106.

\bibitem[{Mike and Farmer(2008)}]{Mike08}
Mike, S., Farmer, J.~D., 2008. An empirical behavioural model of liquidity and
  volatility. Jourmnal of Economic Dynamic and Control 32, 2000.

\bibitem[{O'Hara(1997)}]{OHara97}
O'Hara, M., 1997. Market Microstructure Theory. Blackwell Publishing, Malden,
  USA.

\bibitem[{Paul and Baschnagel(1999)}]{Paul99}
Paul, W., Baschnagel, J., 1999. Stochastic Process from Physics to Finance.
  Springer, Berlin.

\bibitem[{Peng et~al.(1994)Peng, Buldyrev, Havlin, Simons, Stanley, and
  Goldberger}]{Peng93}
Peng, C.-K., Buldyrev, S.~V., Havlin, S., Simons, M., Stanley, H.~E.,
  Goldberger, A.~L., Feb 1994. Mosaic organization of {DNA} nucleotides.
  Physical Review E 49~(2), 1685--1689.

\bibitem[{Plerou et~al.(2005)Plerou, Gopikrishnan, and Stanley}]{Plerou05}
Plerou, V., Gopikrishnan, P., Stanley, H.-E., 2005. Quantifying fluctuations in
  market liquidity: Analysis of the bid ask spread. Physical Review E 71,
  046131.

\bibitem[{Raberto et~al.(2001)Raberto, Cinotti, Focardi, and
  Marchesi}]{Raberto01}
Raberto, M., Cinotti, S., Focardi, S.~M., Marchesi, M., 2001. Agent-based
  simulation of a financial market. Physica A 299, 319.

\bibitem[{Slanina(2008)}]{Slanina08}
Slanina, F., 2008. Critical comparison of several order-book models for stock
  market fluctuations. preprint: physics/0801.0631.

\bibitem[{Smith et~al.(2003)Smith, Farmer, Gillemot, and
  Krishnamurthy}]{Smith03}
Smith, E., Farmer, J.~D., Gillemot, L., Krishnamurthy, S., 2003. Statistical
  theory of the continuous double auction. Quantitative Finance 3(6), 481 --
  514.

\bibitem[{Sornette(2004)}]{Sornette04b}
Sornette, D., 2004. Why stock market crash. Princeton University Press,
  Princeton and Oxford.

\bibitem[{Voit(2005)}]{Voit05}
Voit, J., 2005. The Statistical Mechanics of Financial Markets. Spriger-Verlag,
  Berlin.

\bibitem[{Weber and Rosenow(2005)}]{Weber05}
Weber, P., Rosenow, B., 2005. Order book approach to price impact. Quantitative
  Finance 5(4), 357.

\bibitem[{Weber and Rosenow(2006)}]{Weber06}
Weber, P., Rosenow, B., 2006. Large stock price changes: volume or liquidity?
  Quantitative Finance 6(1), 7.

\bibitem[{Zaccaria et~al.(2010)Zaccaria, Cristelli, Alfi, Ciulla, and
  Pietronero}]{Zaccaria09}
Zaccaria, A., Cristelli, M., Alfi, V., Ciulla, F., Pietronero, L., 2010. The
  asymmetric statistics of the order book: the role of discretness and
  non-uniform limit order deposition. Phys. Rev. E 81, 066101.

\end{thebibliography}

\end{document}